\documentclass[aps,prl,showpacs,twocolumn,amsmath,amssymb]{revtex4}
\usepackage[english]{babel}
\usepackage{latexsym}
\usepackage{graphics}
\usepackage{subfigure}
\usepackage{epsfig}

\begin{document}

\title{Mode coupling control in a resonant device~: application to solid-state ring lasers}

\author{Sylvain~Schwartz$^{1,2}$, Gilles~Feugnet$^1$, Philippe~Bouyer$^2$, Evguenii~Lariontsev$^3$, Alain~Aspect$^2$ and Jean-Paul~Pocholle$^1$}

\affiliation{$^1$Thales Research and Technology France, RD 128, F-91767 Palaiseau Cedex, France\\
$^2$Laboratoire Charles Fabry de l'Institut d'Optique, UMR8501 du CNRS, Centre scientifique d'Orsay B\^at. 503, 91403 Orsay Cedex, France\\
$^3$Lomonosov State University, Moscow, 119992 Russia}

\date{\today}


\email{sylvain.schwartz@thalesgroup.com}

\begin{abstract}
A theoretical and experimental investigation of the effects of mode coupling in a resonant macroscopic quantum device is achieved in the case of a ring
laser. In particular, we show both analytically and experimentally that such a device can be used as a rotation sensor provided the effects of mode
coupling are controlled, for example through the use of an additional coupling. A possible generalization of this example to the case of another
resonant macroscopic quantum device is discussed.
\end{abstract}

\maketitle

Devices using macroscopic quantum effects \cite{Bardeen} and their associated phenomenon of interference to detect rotations can be arbitrarily divided
into two classes \cite{Bretenaker}, namely non-resonant and resonant devices. In devices of the first class, the rotation-induced phase shift is
detected by looking at the displacement of an interference pattern. Among such devices are the fiber optic gyroscope \cite{Lefevre}, the gyromagnetic
gyroscope \cite{Vitale}, the superfluid gyrometer \cite{Varoquaux} and the atomic interferometer \cite{Kasevich}. In devices of the second class, the
rotation is detected through a beat signal. Two examples of resonant and potentially rotation-sensitive devices are the ring laser \cite{Davis} and the
superfluid (e.g. liquid helium or Bose-Einstein condensed gas) in a ring container \cite{Smerzi}. In these two examples, it has been shown
\cite{Leggett,Siegman} that a non-linear mode coupling effect plays a crucial role in the system dynamics, usually preventing it from operating in a
rotation sensitive regime, unless an additional adequate coupling is set.

For example, in the case of the ring laser, the bidirectional emission regime, required for rotation sensing, can be inhibited because the
counter-propagating modes share the same gain medium and can be subject to mode competition. This problem is usually circumvented by choosing for the
gain medium a two isotopes mixture of helium and neon and by tuning the laser emission frequency out of resonance with the atoms at rest. Provided the
detuning value is bigger than the atomic natural linewidth, the gain medium can then be considered as being inhomogeneously broadened and coexistence
of the counter-propagating modes occurs \cite{aronowitz}. In the case of solid-state ring lasers, the gain medium is homogeneously broadened, leading
to a strongly coupled situation resulting in laser emission in only one direction \cite{Siegman}.

We report in this Letter theoretical and experimental investigation of mode coupling control in a solid-state (Nd:YAG) ring laser. The main natural
sources of coupling between the counter-propagating modes are identified, and their role in the laser dynamics is discussed. An additional coupling
source is introduced in order to ensure the coexistence of the counter-propagating modes. A condition for rotation sensing is then analytically
derived, and an experimental confirmation of this theoretical investigation is reported. It is eventually pointed out that the two-level toy model
developed in \cite{Leggett} to describe a superfluid placed in a rotating ring container leads to a similar rotation sensitive operation condition,
opening the way to a possible generalization of this investigation to the case of other resonant macroscopic quantum devices.

The laser equations are derived in the framework of Maxwell-Bloch theory using abiabatic elimination of the polarization \cite{classB}. We assume a
single identical mode in each direction of propagation and neglect transverse effects. The total field inside the cavity is taken as the sum of two
counter-propagating waves~:
\begin{equation}
E(x,t) = \textrm{Re} \left\{ \sum_{p=1}^2 \tilde{E}_{p} (t) e^{i(\omega t + \varepsilon_{p} kx)} \right\} \;, \nonumber
\end{equation}
where $\varepsilon_p=(-1)^p$ and $k=2\pi/\lambda$ is the spatial frequency of the emitted modes associated with the longitudinal coordinate $x$. Using
the slowly-varying enveloppe approximation, the following equations are obtained~\cite{Khanin}~:
\begin{eqnarray}
\frac{\textrm{d}\tilde{E}_{1,2}}{\textrm{d}t}& = & -\frac{\gamma_{1,2}}{2}
\tilde{E}_{1,2}+i\frac{\tilde{m}_{1,2}}{2}\tilde{E}_{2,1} + i \varepsilon_{1,2} \frac{\Omega}{2}\tilde{E}_{1,2} \label{MaxBloch}\\
 & & \!\!\!\!\!\!\!\!\!\!\!\!\!\!\!\!\!\!\!\!\!\!\!\!+\frac{\sigma (1-i \delta)}{2T} \left( \tilde{E}_{1,2}
\int_0^l N \textrm{d}x+ \tilde{E}_{2,1}\int_0^lN{e}^{-2i \varepsilon_{1,2} kx}\textrm{d} x \right) \;, \nonumber \\
 & & \!\!\!\!\!\!\!\!\!\!\!\!\!\!\!\!\!\!\!\!\!\!\!\!\!\!\!\!
  \frac{\partial N}{\partial t} = W_\textrm{th}(1+\eta)-\frac{N}{T_1}-\frac{aN |E(x,t) |^2 }{T_1} \;, \label{MaxBloch2}
\end{eqnarray}
where $\gamma_{1,2}$ are the cavity mode losses per time unit (we will first assume $\gamma_1=\gamma_2 =\gamma$), $\sigma$ the stimulated emission
cross section, $T$ the cavity round trip time, $l$ the length of the gain medium, $N(x,t)$ the population inversion density function, $\eta$ the excess
of pump power above the threshold value $W_\textrm{th}$, $T_1$ the population inversion relaxation time and $a$ the saturation parameter. The
rotation-induced angular eigenfrequency difference $\Omega$ between the counter-propagating modes is linked to the angular velocity $\dot{\theta}$ by
the Sagnac formula \cite{Sagnacarticle}~:
\begin{equation} \label{Sagnac}
\Omega = \frac{8 \pi A}{\lambda c T} \dot{\theta} \;,
\end{equation}
where $A$ is the area enclosed by the ring cavity and $c$ the speed of light in vacuum. The parameter $S=4A/(\lambda c T)$ is known as the scale factor
of the cavity. The detuning of the cavity from the center of the gain line is defined as $\delta = (\omega_c - \omega_{ab})/\gamma_{ab}$, where
$\omega_c = k c$ is the resonance frequency of the modes and $\omega_{ab}$ and $\gamma_{ab}$ are respectively the position of the center and the width
of the gain line. For solid-state gain media, $\delta$ is usually smaller than $10^{-2}$. Its effects (among which is the dispersion of the refractive
index) will therefore be neglected in our analysis.

Because of the scattering of light induced by the mirrors and by the amplifying crystal, a fraction of the power of each mode is injected back into the
other, resulting in mutual coupling. This effect has been taken into account phenomenologically in equation (\ref{MaxBloch}) using the backscattering
coefficients $\tilde{m}_{1,2}=me^{ i \varepsilon_{1,2} \theta_{1,2}}$. It can cause phase synchronization and intensity stabilization of the
counter-propagating modes. The strength of this coupling decreases when the speed of rotation increases, as the difference between the eigenfrequencies
of the counter-propagating modes becomes more and more important. Note that even if the backscattering coefficient is usually small (i.e. $m\ll \gamma
\eta$), this coupling still has to be taken into account for a correct description of the modes dynamics.

Another source of coupling is caused by the establishment of a population inversion grating in the gain medium, created by the light pattern resulting
from the interference between the counter-propagating waves. This coupling corresponds in equation (\ref{MaxBloch}) to the term proportional to the
spatial Fourier transform of $N$ at the order $2k$, and can be interpreted as resulting from backward diffraction on the grating. When the speed of
rotation increases, the contrast of the grating (hence the coupling strength) decreases because of the movement of the light interference pattern and
of the inertia of the gain medium. More precisely, under the conditions~:
\begin{equation} \label{conditions}
|\Omega| \gg \sqrt{\frac{\gamma \eta}{T_1}} \, , \, m \qquad \textrm{and} \qquad \eta \ll 1 \;,
\end{equation}
a perturbation method applied to equations (\ref{MaxBloch}) and (\ref{MaxBloch2}) shows that the amplitudes of the counter-propagating modes are
coupled through the gain-induced effective coupling coefficient $\tilde{N}$ given by~:
\begin{equation} \label{eqN}
2\tilde{N}=\frac{\gamma \eta}{1+\Omega^2 T_1^2} \;.
\end{equation}
The presence of the grating hinders the coexistence of the two counter-propagating modes because it results in the cross-saturation parameter being
bigger than the self-saturation parameter which leads, in the absence of other coupling source, to single mode (unidirectional) operation
\cite{Siegman}.

In order to counteract this effect and to achieve rotation sensitive operation, a stabilizing additional source of coupling is provided to the system.
This is done by producing losses which depend on the intensity difference between the counter-propagating modes according to the following law~:
\begin{equation}\label{pertesdiff}
\gamma_{1,2}=\gamma - \varepsilon_{1,2} K \left(a|\tilde{E}_1|^2-a|\tilde{E}_2|^2\right) \;,
\end{equation}
where $K$ is chosen to be positive so that the mode with the higher intensity gets the higher losses. The associated effective coupling coefficient is
on the order of $K \eta$. A concrete manner of generating such losses will be detailed further in this Letter.

\begin{figure}
\begin{center}
\includegraphics[scale=0.6]{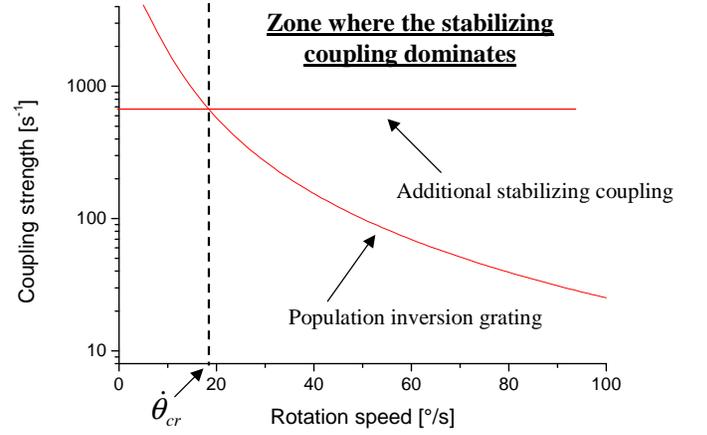}
\end{center}
\caption{Effective coupling coefficients $\tilde{N}$ and $\eta K$ as a function of the rotation speed $\dot{\theta}$ (logarithmic vertical scale). The
two ranges of rotation speed are delimited by the value $\dot{\theta}_{cr}$, given by equation (\ref{tethacritique}). Below this value the coupling due
to the population inversion grating dominates, while above this value it is the additional stabilizing coupling that dominates.}\label{couplings}
\end{figure}

Using typical values of the parameters, we plotted $\tilde{N}$ and $K \eta$ as a function of $\dot{\theta}$ on figure \ref{couplings}. Two ranges of
rotation speed rise, delimited by the following critical value~:
\begin{equation} \label{tethacritique}
\dot{\theta}_{cr} = \frac{1}{2\pi S T_1} \sqrt{\frac{\gamma}{K}-1} \;.
\end{equation}
In the zone where $\dot{\theta} < \dot{\theta}_{cr}$ the coupling generated by the population inversion grating dominates, while in the zone where
$\dot{\theta} > \dot{\theta}_{cr}$ it is the additional stabilizing coupling that dominates. This latter zone turns out to be the zone of rotation
sensitive operation, as we will see further.

To proceed, equations~(\ref{MaxBloch}) and (\ref{MaxBloch2}) have been solved under the conditions (\ref{conditions}). We looked for a solution
corresponding to the beat regime, i.e. obeying the following conditions~:
\begin{equation} \label{selfcons}
\left|\frac{|\tilde{E}_1|^2-|\tilde{E}_2|^2}{|\tilde{E}_1|^2+|\tilde{E}_2|^2}\right| \ll 1 \qquad \textrm{and} \qquad |\dot{\Phi}-\Omega| \ll |\Omega| \;,
\end{equation}
where $\Phi$ is the difference between the arguments of $\tilde{E}_1$ and $\tilde{E}_2$. We obtained the following expression for the relative intensity
difference in the beat regime~:
\begin{equation} \label{difference}
\frac{|\tilde{E}_1|^2-|\tilde{E}_2|^2}{|\tilde{E}_1|^2+|\tilde{E}_2|^2} = \frac{m^2\sin (\theta_1- \theta_2)}{8 \Omega ( \tilde{N} + \eta K)}.
\end{equation}
The existence of the beat regime is subject to the self-consistency condition (\ref{selfcons}) for high rotation speeds. As can be seen on
equation~(\ref{difference}), this depends both on $\tilde{N}$ and on $\eta K$.

If we first consider the absence of additional coupling ($K=0$), we see on equation (\ref{difference}) using expression (\ref{eqN}) that the relative
intensity difference diverges linearly with $\Omega$. This is in contradiction with condition (\ref{selfcons}) and leads to the conclusion that the
beat regime does not exist in this case. Instead, the laser might turn to unidirectional operation \cite{Lariontsev4}.

In the case where the additional coupling is present ($K\neq 0$), the relative intensity difference goes to zero like $1/\Omega$, satisfying the
self-consistency condition (\ref{selfcons}). It can then be shown that the beat regime is stable if~:
\begin{equation} \label{condition}
\eta K > \tilde{N} \;.
\end{equation}
Inequality (\ref{condition}) is equivalent to $\dot{\theta}>\dot{\theta}_{cr}$, which means that the second zone of figure \ref{couplings} is indeed
the zone of rotation sensing operation. It is remarkable that even for very small (positive non-zero) values of $K$ the beat regime is stable for
sufficiently high rotation speeds.

\begin{figure}
\begin{center}
\includegraphics[scale=0.75]{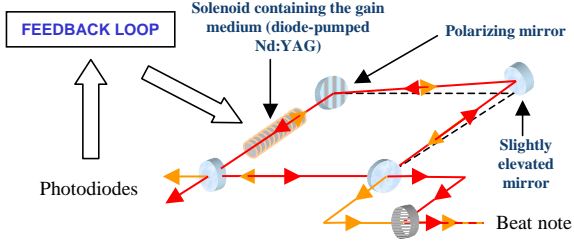}
\end{center}
\caption{Schematic representation of the experimental setup. We used a non-planar ring laser cavity of about 25~cm of perimeter, enclosing a surface
of about 34 cm$^2$ and with a 2 cm-long Nd:YAG rod. The skew rhombus angle is on the order of $10^{-2}$~rad. The whole device is placed on a turntable.}
\label{figure1}
\end{figure}

We now report experimental achievement of the device described above. The stabilizing additional coupling is provided to the CW diode-pumped Nd-YAG
ring laser cavity using the setup of figure~\ref{figure1}. This configuration, which is inspired from the one used in ring lasers in order to achieve
unidirectional single-mode operation \cite{Biraben}, is based on the combination of three polarization-related effects. The first effect is a
reciprocal rotation of the polarization plane by the use of a slightly non planar cavity. The rotation angle $\alpha$ depends on the geometry of the
cavity. The second effect is a non reciprocal (Faraday) rotation of the polarization plane, produced by a solenoid placed around the Nd-YAG rod. The
rotation angle $\beta$ is proportional to the current flowing through the solenoid. The third effect is a polarizing effect, achieved by replacing one
of the cavity mirrors with a polarizing mirror. The differential losses are then to the first order equal to $\gamma_1-\gamma_2 = 4\alpha \beta / T$.
The light intensities of the two counter-propagating modes are monitored by two photodiodes, and the value of $\beta$ is kept proportional to the
difference between those intensities by the mean of an electronic feedback loop acting on the current inside the solenoid. This results in differential
losses of the form~(\ref{pertesdiff}).

\begin{figure}
\begin{center}
\includegraphics[scale=0.65]{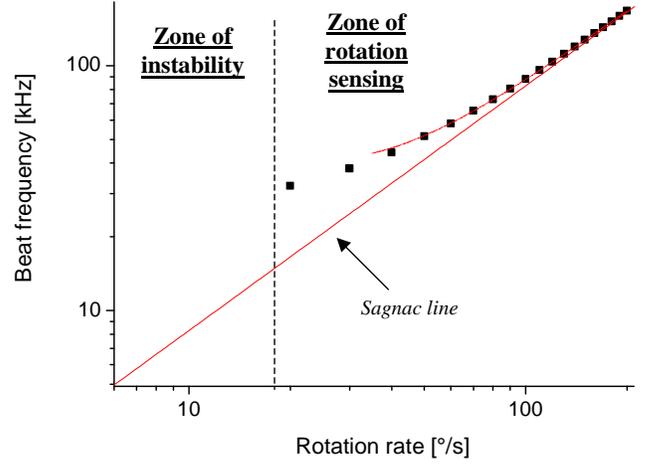}
\end{center}
\caption{Experimental frequency response of the solid state ring laser gyroscope for $\eta \simeq 0,17$ (logarithmic horizontal and vertical scales).
For rotation speeds lower that $\dot{\theta}_{cr} \simeq 19^\circ/s$, the stabilizing coupling is dominated by the effects of the population inversion
grating and no beat frequency is measured. From the asymptotic linear dependance, we get an experimental value for the scale factor
$S$~=~0,83~$\pm$~0,01~kHz/($^\circ$/s). This is in perfect agreement with the Sagnac formula (\ref{Sagnac}) for the dimensions and the emission
wavelength of our experimental setup.}\label{figure2}
\end{figure}

The two ranges of figure \ref{couplings} have been identified experimentally, with a measured value of $\dot{\theta}_{cr} \simeq 19 ^\circ /s$. Below
this critical rotation speed, we observe instabilities of the intensities of the modes, and no beat signal. Above $\dot{\theta}_{cr}$, the intensities
of the two modes are stable and similar in magnitude and a beat signal is observed. The measured frequency of this beat signal as a function of the
rotation speed is reported on figure~\ref{figure2}.

\begin{figure}
\begin{center}
\includegraphics[scale=0.6]{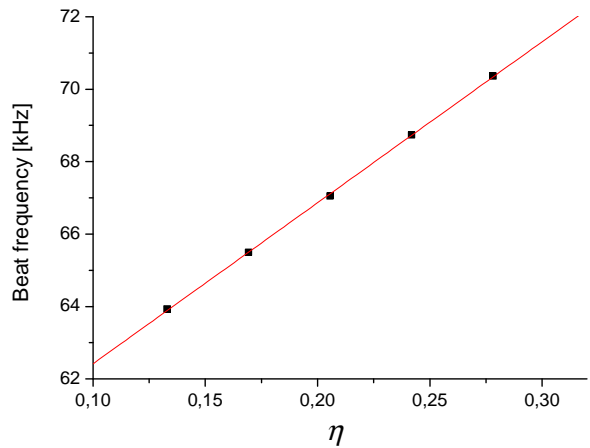}
\end{center}
\caption{Beat frequency as a function of the pumping rate for $\dot{\theta}=70^\circ/s$ (error bars are smaller than 100 Hz). The observed lienar
shift, in agreement with equation (\ref{beat}), is a direct manifestation of the mode coupling due to the population inversion grating in the
solid-state ring laser.}\label{figure3}
\end{figure}

This frequency response curve matches the Sagnac line for high values of the rotation speed. For slower rotations (but still above
$\dot{\theta}_{cr}$), we observe a deviation from the Sagnac line. This deviation is partly due to the linear coupling induced by backscattering on the
mirrors and on the Nd-YAG crystal (Adler pulling) and partly due to the coupling induced by the population inversion grating (this last phenomenon can
be interpreted as resulting from Doppler shift on the moving population inversion grating). The beat frequency $<\dot{\Phi}>$ (where $<>$ stands for
time averaging) can be expressed analytically under the condition (\ref{selfcons}), leading to the following formula~:
\begin{equation} \label{beat}
<\dot{\Phi}> = \Omega + \frac{m^2 \cos(\theta_1-\theta_2)}{2 \Omega}+ \frac{\gamma \eta}{2 \Omega T_1} \;.
\end{equation}
This expression is in good agreement with our experimental data for $\dot{\theta}\gtrsim 50^\circ/s$ (see figure \ref{figure2}). The linear dependance
of $<\dot{\Phi}>$ on the pump power (represented by the parameter $\eta$) for a fixed rotation speed has been checked experimentally. The result is
given for $\dot{\theta}=70^\circ/s$ on figure~\ref{figure3}. This is a direct manifestation of the coupling between the counter-propagating modes
induced by the population inversion grating in a solid-state ring laser.

Before concluding, it is worth noting that the physical system composed of a superfluid in a ring container has some similarities with the system
described in this Letter, at least in the limit of the two-level toy model of \cite{Leggett}. In this model, the condition for the system to be
rotation sensitive reads~:
\begin{equation} \label{superfluid}
V_0>g \;,
\end{equation}
$V_0$ being the asymmetry energy and $g$ the mean (repulsive) interaction energy per particle in the $s$-wave state.
Condition (\ref{superfluid}) reflects the fact that a change in the quantum of circulation around the ring, which is a
signature of rotation, is more difficult when repulsive interactions are stronger and is easier when the trap is
asymmetric. Condition (\ref{superfluid}) has to be compared with condition (\ref{condition}). In both cases, these
conditions express the fact that the system is rotation-sensitive provided that a 'good' additional coupling is stronger
than a 'bad' non-linear coupling. We expect that achieving condition (\ref{superfluid}) in a physical system will lead to
new atomic sensors concepts in the same way as achieving condition (\ref{condition}) in an optical system allows the
realization of a solid-state ring laser gyroscope.

In conclusion, we have shown that the solid state ring laser provides a good illustration of mode coupling control in a resonant macroscopic quantum
device. In particular, rotation sensing has been achieved by the use of a stabilizing mode coupling effect that counteracts the destabilizing effect of
the population inversion grating. The simple theoretical model we used is in good agreement with the reported experimental results. This work opens new
perspective for the realization of solid state laser gyrometers. In addition, it shows the importance of interplay between mode couplings in a system
with periodic boundary conditions. The same concept can be applied in more complex systems like toroidal superfluids, a field in which an important
experimental effort is being made at the moment \cite{StamperK}.

The authors acknowledge constant support from Thales Aerospace Division, and thank A. Garnache and S. Richard for fruitful
discussion and J. Colineau and J. Antoine for making the electronic feedback circuit. S. S. thanks J. Gaebler and S.
Demoustier for rereading the manuscript. E. L. thanks the French Russian GDRE Lasers and Optoelectronics for supporting
his stay in France as an invited visiting research fellow.


\begin{references}

\bibitem{Bardeen} J. Bardeen, Phys. Today \textbf{43}, No. 12, 25 (1990)

\bibitem{Bretenaker} F. Bretenaker et al., Phys. Rev. Lett. \textbf{69}, No. 6, 909 (1992)

\bibitem{Lefevre} H. Arditty and H. Lef\`evre, Opt. Lett. \textbf{6}, No. 8, 401 (1981)

\bibitem{Vitale} S. Vitale et al., Phys. Rev. B \textbf{39}, No. 16, 11993 (1989)

\bibitem{Varoquaux} O. Avenel et al., Phys. Rev. Lett. \textbf{78}, No. 19, 3602 (1997)

\bibitem{Kasevich} T. L. Gustavson et al., Phys. Rev. Lett. \textbf{78}, 2046 (1997)

\bibitem{Davis} W. Macek and D. Davis, Applied Phys. Lett. \textbf{2}, No. 3, 67 (1963)

\bibitem{Smerzi} M. Benakli et al., Europhys. Lett. \textbf{46}, No. 3, 275 (1999)

\bibitem{Leggett} A. Leggett, Rev. Mod. Phys. \textbf{73}, 307 (2001)

\bibitem{Siegman} A. Siegman, \textit{Lasers}, University Science Books (1986)

\bibitem{aronowitz} F. Aronowitz in \textit{Laser applications}, Academic Press, 133 (1971)

\bibitem{classB} H. Zeghlache et al., Phys. Rev. A \textbf{37}, No. 2, 470 (1988)

\bibitem{Khanin} P. Khandokhin and Y. Khanin, J. Opt. Soc. Am. B \textbf{2}, No. 1, 226 (1985)

\bibitem{Sagnacarticle} G. Sagnac, C.R. Acad. Sci. \textbf{157}, 708 (1913)

\bibitem{Lariontsev4} N. Kravtsov, E. Lariontsev and A. Shelaev, Laser Phys. \textbf{3}, No. 1, 21 (1993)

\bibitem{Biraben} F. Biraben, Optics Comm. \textbf{29}, No. 3, 353 (1979)

\bibitem{StamperK} S. Gupta et al., Phys. Rev. Lett. \textbf{95}, 143201 (2005)

\end{references}
\end{document}